\documentstyle[preprint,tighten,prc,aps,epsfig]{revtex}
\newcommand{\dis}{\displaystyle}
\newcommand{\be}{\begin{eqnarray}}
\newcommand{\ee}{\end{eqnarray}}
\def\bra#1{\textstyle{\left\langle \, #1 \, \right\vert \>}}
\def\ket#1{\textstyle{\> \left\vert \>\! #1 \>\! \right\rangle}}

\begin{document}

\normalsize

\title{Production of $\eta$ mesons in nucleon-nucleon collisions}

\author{V. Baru$^{a,b}$, A.M. Gasparyan$^{a,b}$, J. Haidenbauer$^{a}$,
C. Hanhart$^{a}$, A.E. Kudryavtsev$^{b}$, and J. Speth$^{a}$}

\address{
$^a$Institut f\"{u}r Kernphysik, Forschungszentrum J\"{u}lich
GmbH,\\ D--52425 J\"{u}lich, Germany \\
$^b$Institute of Theoretical and Experimental Physics, \\
117259, B. Cheremushkinskaya 25, 
Moscow, Russia }

\maketitle 

\begin{abstract}
A microscopic calculation of near-threshold
$\eta$-meson production in the reaction $NN \rightarrow NN\eta$ 
is presented. It is assumed that the $\eta$ meson is produced 
via direct emission and via elementary rescattering processes 
$M N \rightarrow \eta N$ of various mesons
$M = \pi, \ \rho$, etc. As a novel feature 
the amplitudes for the latter production mechanism are taken from a 
multi-channel meson-exchange model of the $\pi N$ system 
developed by the J\"ulich group which contains explicitly the 
channels $\pi N$, $\rho N$, $\eta N$, $\sigma N$, and $\pi \Delta$. 
Furthermore, effects of the $NN$ interaction in the final 
as well as in the initial state are taken into account 
microscopically. Our results are compared with 
recent data from the COSY and CELSIUS accelerator facilities.
Reasonable agreement with available near-threshold cross section
data for the reactions $pp\to pp\eta$, $pn\to pn\eta$, 
and $pn\to d\eta$ is found. 
\end{abstract}

\vskip 1cm \noindent


\vskip 2cm

\hfill{FZJ--IKP--TH--2002--16}

\newpage

\section{Introduction}

Meson production in nucleon-nucleon ($NN$) and
nucleon-nucleus collisions is a rather attractive
research field in hadron physics at intermediate energies.
The revived interest in such reactions is closely connected to
recent successful developments in the accelerator technology which 
enabled experimentalists  
to perform rather accurate measurements on the production of 
$\pi$, $\eta$, $\eta'$ and other mesons in nucleon-nucleon collisions in the 
near-threshold region \cite{Haidrev,COSYrev,Wilrev}.  
These data in combination with the theoretical analysis open the unique 
possibility to study the mechanism of the 
production of different mesons as well as to produce information
on resonance properties, baryon--baryon and meson--baryon interactions, 
short range properties of the $NN$ interaction etc..

Among those reactions the production of the $\eta$ meson  
is certainly rather interesting. E.g., the $\eta$  
is the only other non-strange pseudo Goldstone boson
existing and thus a close though heavier relative of the pion.   
Also it is expected that $\eta$ production near threshold is closely 
linked with the properties of the $S_{11}$ $N^*$(1535) resonance.
Finally, and may be most importantly there is already a wealth of
rather accurate data on $\eta$ meson production near threshold 
providing a sensible testing ground for model calculations. 
Apart from total cross sections for the 
reaction $pp\to pp \eta$ \cite{Berg,Chia,Ups,Hibou,COSY,COSY1}, 
there are also some data on the total cross sections of the 
reactions $pn\to pn \eta$ \cite{Calenpneta} and 
$pn\to d \eta$ \cite{Upspndeta1,Upspndeta2}. Furthermore, 
there are data on differential cross sections and 
invariant-mass spectra \cite{COSY1,Calendiff,COSYTOF,Moskal02} 
and even anlyzing powers for the reaction $p p\to pp \eta$ \cite{Wint}. 

The present work is devoted to the study of 
the production of $\eta$ mesons in $NN$ collisions.
The goal is a combined theoretical analysis of all measured channels 
of the reaction $NN\to NN\eta$, namely $pp \to pp \eta$,
$pn \to pn \eta$ and $pn \to d\eta$,
by taking into account consistently the interaction 
between the nucleons in the final as well as in the initial
state and by utilizing a microscopic model of meson-nucleon ($MN$)  
scattering \cite{Schutz,Olipap} for the description of the 
$\eta$-meson production process. 
We focus on the description of $\eta$-meson production
in the near-threshold region, i.e. for excess energies $Q$ up to around 50 MeV. 
Therefore, we will restrict our investigation to S-wave contributions. 
Recent data on differential cross sections of the reaction 
$pp\to pp\eta$ \cite{COSYTOF} suggest that in this energy region 
the production of $\eta$ mesons still takes place mainly in S-waves.

\begin{center}
\begin{figure}
\epsfig{file=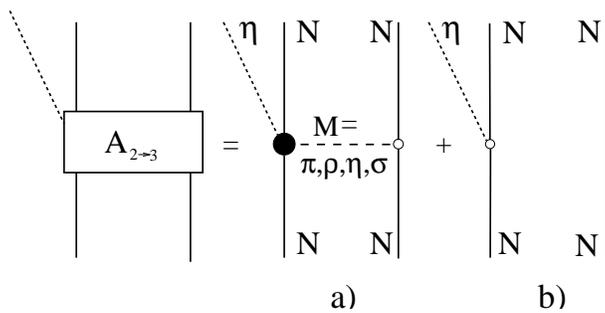, width=8cm}
\vspace*{0.5cm}
\caption{Production mechanisms for the reaction $NN \to NN\eta$ 
taken into account in our model: (a) $\eta$ production via $MN \to \eta N$ 
rescattering; (b) direct $\eta$ production.}
\label{diag}
\end{figure}
\end{center}

In spite of the abundance of theoretical investigations 
on $\eta$-meson production 
\cite{Lag,Moal,Ged,Santra,Ber,Bat,Wilppeta,Wilnew,Vett,Kochelev,Dillig,Pena,Grishina,Nakayama}
the leading production mechanism is still not identified.
There is general consent in the literature that $\eta$-meson production
is dominated by re-scattering processes as depicted in Fig.~\ref{diag}a,
with a variety of intermediate meson exchanges. 
However, it is not clear which
one of the possible meson exchanges plays the most important role.
For instance, in Refs. \cite{Ged,Santra,Wilppeta,Wilnew} 
it is suggested that the dominant contributions should come from the   
$\rho$ meson, whereas in \cite{Vett,Pena} it 
is found to be of minor importance compared to other contributions 
and in \cite{Bat} that particular contribution is not even considered.
Recently Nakayama et al. \cite{Nakayama} compared the two scenarios of a 
$\pi$- and a $\rho$-exchange dominance and concluded that both these 
scenarios can describe the cross section data for the various
$NN\to NN\eta$ channels equally well.


Apart from the choice of the exchanged mesons there are also 
differences in the
construction of the $MN\to \eta N$ transition amplitude. Many of the 
models \cite{Lag,Moal,Ged,Santra,Wilppeta,Vett,Nakayama} are based 
on the calculation of tree level re-scattering diagrams where the  
$MN\to \eta N$ transition amplitude is simply parametrized by the 
$S_{11}$ $N^*(1535)$ resonance. 
Background contributions, e.g., through $t$--channel meson exchanges are 
either omitted or simply added on top of the resonance contribution.
In such approaches the relative phases between different 
meson exchanges have to be fixed by hand, which naturally introduces large
uncertainties. 
As a consequence quite contradictory prescriptions have been adopted in
the literature. E.g., in Ref. \cite{Ged} it was tried to minimize 
interferences between the various production mechanisms. Assuming that
the $\eta$'s are produced only via $N^*$ excitation led to $\pi$ and
$\rho$ exchange contributions that are orthogonal to each other and,
therefore, don't interfere at all. On the other hand, 
in a recent investigation by F\"aldt and Wilkin \cite{Wilnew} 
interferences are sort of maximized. Those authors argue that
only a strong interference of the $\rho$ exchange amplitude with the 
$\pi$ exchange contribution allows  
to achieve a consistent description of the experimentally observed 
cross sections for the reactions $pp\to pp\eta$ and ${pn\to pn\eta}$.

In the present investigation we employ $MN \to \eta N$ amplitudes generated
from a microscopic (meson-exchange) model of the $\pi N$ system. 
It is a coupled-channels model that contains 5 channels, namely 
$\pi N$, $\eta N$, $\rho N$, $\sigma N$, and $\pi \Delta$ \cite{Olipap}. 
The interactions in and between the various
channels are derived in the meson-exchange picture starting
out from effective chiral Lagrangians. The model includes the $N^* (1535)$ resonance
as essential contribution but in addition also various ($t$--channel) meson
and ($u$--channel) baryon exchange diagrams. 
The parameters of the model are fixed by requiring a simultaneous description of 
the $\pi N$ phase shifts and inelasticities as well as of the transition cross 
sections for the reactions $\pi N\to \eta N$ and 
$\pi N\to \rho N$ over an energy range that extends well beyond the $\eta N$
threshold. The reaction amplitudes are obtained from solving a 
coupled-channels relativistic scattering equation of Lippmann-Schwinger type.
Clearly, in such a model not only the $\pi N\to \eta N$ amplitude is
determined by empirical data, but also the transition amplitudes involving 
heavier mesons are to a large extend constrained by the phase shifts and 
inelasticity parameters of $\pi N$ scattering. This is also true for the 
relative phases between the various amplitudes.

Another uncertainty in investigations of the reaction $NN\to NN\eta$ 
is caused by the treatment of the $NN$ FSI and ISI effects.
In many papers concerned with meson production FSI effects are 
taken into account approximately by applying methods which are rather 
questionable if one wants to obtain absolute predictions
for the production cross sections (see, e.g., Refs. 
\cite{Lag,Moal,Ged,Ber,Vett} for 
the case of $\eta$-production), as was pointed out recently in Refs.
\cite{H-N,BarFSI}.  The situation with the ISI is even
less satisfying. Indeed, its effect is simply omitted in most studies. 

In our investigation we take the FSI and ISI in the $NN$ system
fully into account. In particular, we employ an $NN$ model that
reproduces the relevant $NN$ phase shifts reasonably well up to energies 
around the $\eta$ production threshold and, therefore, allows a consistent 
description of the FSI and ISI effects. 

Let us mention here that, at the present stage, 
the J\"ulich $MN$ model does not include the $\omega N$ channel. Thus, we 
cannot take into account the $\omega$-rescattering contribution to 
the reaction $NN\to NN\eta$. Anyhow, we want to emphasize 
that most model studies indicate that its contribution
should be small \cite{Moal,Ged,Wilnew,Nakayama}.

The paper is structured in the following way:  
In Sect. II we describe the ingredients that enter into our model calculation 
of $\eta$ production in nucleon-nucleon collisions. In particular,
we present some details about the meson-nucleon amplitudes that are
utilized for constructing the elementary $\eta$ meson production operator
and we specify the relevant vertex parameters. We also provide a brief
description of the $NN$ model that is used for the treatment of the
ISI and FSI, and we supply its prediction for those energies and partial waves 
in the initial $NN$ state, relevant for $\eta$ production near threshold. 
Our results are presented and discussed in Sect. III. We show our
predictions for various charge channels, i.e. 
$pp \to pp\eta$, $pn \to pn\eta$, and $pn \to d\eta$, and
compare them with available data on total cross sections. 
We also study the sensitivity of our results to variations in the
elementary ($\rho N \to \eta N$) production amplitude and to 
differences in the $NN$ interaction in the final state. 
Finally, we analyse the role played by the various production mechanisms
included in our model. 
A summary and some concluding remarks are given in Sect. IV.

\section{Model calculation of the reaction $NN\to NN\eta$}

\begin{figure}
\begin{center}
\epsfig{file=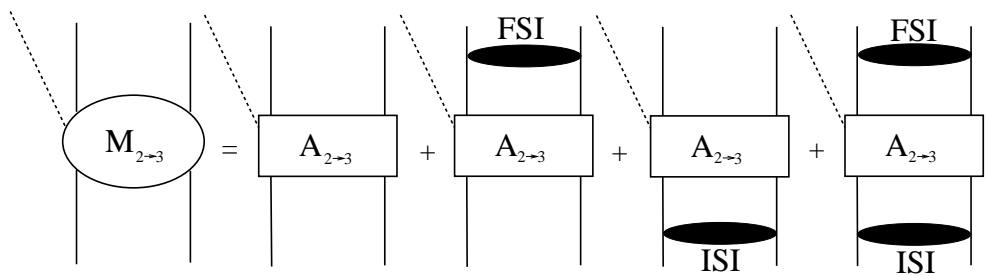,width=13cm}
\vspace*{0.5cm}
\caption{Diagramatic representation of our DWBA calculation.
$A_{2\to 3}$ is the elementary $\eta$-production amplitude. 
The filled ellipses stand for the $NN$ interaction in the
final and initial states.}
\label{Mamp}
\end{center}
\end{figure}

We study $\eta$-meson production in distorted wave Born approximation (DWBA).
This means that the transition amplitude from the initial to the final 
state, $M_{2\to 3}$, is given by the expression 
\be
M_{2\to 3}=\bra{\chi_{f}} (1+T_{NN}^{FSI}G_0) A_{2\to 3} (1+G_0T_{NN}^{ISI})\ket{\chi_{i}}.
\label{M23}
\ee 
where $A_{2\to 3}$ is the (elementary) production operator and 
$T_{NN}^{ISI}$ and $T_{NN}^{FSI}$ are the $NN$ reaction amplitudes in the initial 
and final states. A graphical representation of Eq.~(\ref{M23}) is given in
Fig.~\ref{Mamp}. Note that here we have tacitly assumed that the
final-state effects are dominated by the
interaction in the $NN$ system and that the interaction 
in the $\eta N$ system can be neglected. There is, however, 
experimental evidence \cite{Ups,Upspndeta1,Moskal}
that the $\eta N$ interaction has a significant impact 
on the energy dependence of the cross section.
Under such circumstances it would be, in principle, desirable to include 
the $\eta N$ FSI consistently and to solve 3-body integral equation 
of Faddeev type for the most difficult case
when all particles are in the continuum (see, e.g. Ref. \cite{Schmid} 
and references therein). A few efforts along this line can be already
found in the literature \cite{Ueda,Pena02}, though only for the quasi 
two-body reaction $np\to d\eta$.
On the other hand, the experiments also indicate that the $\eta N$ interaction 
might have an influence only for very small excess energies, i.e. up to 
around 10-15 MeV above the $\eta$-threshold. 
Thus, a sensible investigation of the $\eta$-meson production mechanisms 
should be still feasible without taking into account the $\eta N$ interaction, 
if one focusses on the energy range $10 \le Q\le 50$ MeV, say. 
In any case it is quite reasonable to consider the much more complicated 
3-body effects only after a detailed quantitative understanding of 
$NN$ ISI and FSI effects as well as of the production process has been achieved.
Therefore, in this work we disregard possible effects from the $\eta N$ FSI.


The production amplitude $A_{2\to 3}$ in our model consists of the re-scattering diagrams
with $\pi,\rho,\eta,\sigma$ meson exchanges and the direct $\eta$ production. 
Corresponding diagrams included in our work are shown in  Fig. \ref{diag}.

One of the principal novelties of our investigation is the utilization of a 
realistic microscopic model for the elementary reaction amplitudes 
$MN\to \eta N$, 
namely a coupled-channels model for $\pi N$ scattering 
that has been developed recently by the J\"ulich group \cite{Olipap,Olithes}. 
The important aspect connected with the use of this model is that the
off-shell properties of those amplitudes and also the 
relative phases between different meson contributions are now fixed and (to
a certain degree) constrained by the data on $\pi N$ scattering. 
Another merit of our model study of $\eta$ production in $NN$ collisions
consists in the full and consistent treatment of effects from the $NN$ 
FSI as well as ISI.

As mentioned already we restrict our investigation to the S-wave 
contributions. Recently measured angular distributions in the 
reaction $pp\to pp\eta$ exhibit an isotropic structure at 
the excess energies $Q \ =$ 15 MeV as well as at 
$Q \ =$ 41 MeV \cite{COSYTOF}, thus suggesting that even up to 
$Q \approx$ 50 MeV the production of 
$\eta$ mesons might take place predominantly in S-waves.
We want to point out, however, that the invariant-mass distributions 
reported in the same paper cannot be described by S-waves
plus $pp$ FSI alone \cite{COSYTOF}. This could be an evidence that 
at least at the higher energy P-waves already play a role \cite{Naka2}. 

Restricting ourselves to S-waves
we have only two amplitudes corresponding to the
possible final states $^1S_0s$ (for isospin I=1) and $^3S_1s$ 
(for isospin I=0). The corresponding initial $NN$ states are 
$^3P_0$ and $^1P_1$, respectively. (We use the standard nomenclature
where capital letters denote the $NN$ partial waves and the small
letter labels the orbital angular momentum of the $\eta$ with 
respect to the $NN$ system.)

\subsection{Model of the meson-nucleon ($MN$) interaction}
\label{MN}

In this subsection we describe the structure and results of the $MN$ model 
-- which is one of the main ingredients of our calculation. 
This model was developed in Refs. \cite{Schutz,Olipap,Olithes}. 
The ambitious aim of the model is a simultaneous description of the $\pi N$ phase
shifts and inelasticities as well as of the transition cross sections 
for the reactions $\pi N\to \eta N$ and 
$\pi N\to \rho N$. The reaction amplitudes are obtained from solving
a relativistic coupled-channels scattering equation of Lippmann-Schwinger type.
The model includes 5 channels, namely $\pi N$, $\eta N$, $\rho N$, $\sigma N$, 
and $\pi \Delta$.
All the $MN$ interactions and transition potentials are constructed 
in time-ordered perturbation theory (TOPT) on the basis of effective chiral
Lagrangians. Most of the coupling constants at the meson-baryon-baryon and 
meson-meson-meson vertices included in the model are taken
from other sources \cite{Olipap}.
The cutoff masses in the vertex form factors 
are the only free parameters which were determined by a fit to
the $\pi N$ partial-wave amplitudes. The potentials consist of 
$t$-channel meson exchanges, $s$-channel resonance graphs and 
$u$-channel baryon exchanges (cf. Figs. 2-5 in Ref. \cite{Olipap}).

\begin{figure}[h]
\begin{center}
\epsfig{file=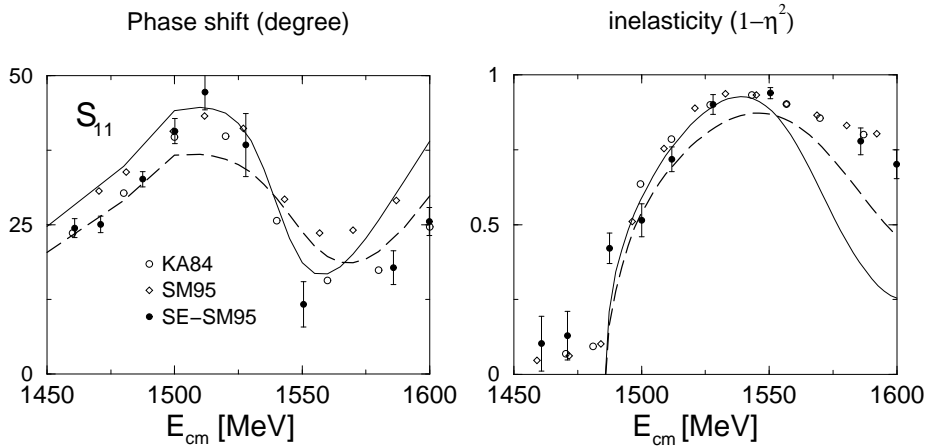, width=12.2cm}
\vspace*{0.5cm}
\caption{The $\pi N$ phase shift and inelasticity $1-\eta^2$ for the 
$S_{11}$ partial wave. The dashed curves represent the results of 
the original $\pi N$ model of Ref. \protect\cite{Olipap,Olithes}
whereas the solid lines are the results of the extended model
that includes the additional diagrams of Fig. \ref{rhodiag}.} 
\label{S11rho}
\end{center}
\end{figure}

The model provides a satisfactory overall description of all $\pi N$ partial 
amplitudes with total angular momentum $J\le 3/2$ \cite{Olipap}.
Specifically, in the $S_{11}$ partial wave, which is {\it the} relevant one for 
the reaction $NN\to NN\eta$ close to threshold,
the agreement with the experimental information 
in a region of about 60 MeV above the $\eta$ threshold 
(the $\eta N$ threshold is at $E_{cm} \approx 1486$ MeV) is quite 
good, as demonstrated in Fig. \ref{S11rho} by the dashed lines.
Therefore this model is a reasonable starting point for our calculation.
The off-shell transition amplitudes of this model are utilized for the
meson rescattering amplitudes $MN\to \eta N$ that enter into the evaluation
of the production operator $A_{2 \to 3}$ 
(cf. the filled circle in Fig. \ref{diag}a). 

\begin{figure}
\begin{center}
\epsfig{file=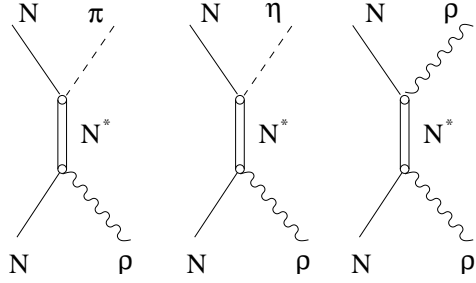,width=6.2cm}
\vspace*{0.5cm}
\caption{Additional diagrams containing the $\rho NN^*$ coupling which are 
included in the extended model.}
\label{rhodiag}
\end{center}
\end{figure}

It should be said, however, that the data on $\pi N$ scattering 
(plus the $\pi N \to \eta N$ and $\pi N \to \rho N$ transition cross sections)
do not fully determine all the transition potentials of the coupled-channels 
model. E.g., in the original model \cite{Olipap} no direct 
$\rho N\to \eta N$ transition potential at all was included, 
because the $\pi N$ results are not too sensitive to such contributions 
and, accordingly, corresponding parameters could not be fixed
unambigously. Possible contributions to $\rho N\to \eta N$ can arise from 
$t$-channel meson-exchange (e.g., $\rho$, $b_1$, etc.), from $N$ exchange 
and, in particular, from the $s$-channel pole diagram involving the
$S_{11}$ $N^*(1535)$ resonance, as shown in Fig. \ref{rhodiag}. 

Note that the original $\pi N$ model \cite{Olipap} still generates a 
$\rho N\to \eta N$ transition amplitude - even in the absense of a
direct $\rho N\to \eta N$ transition potential. 
Such an amplitude arises naturally in a coupled-channels model, e.g. 
via higher order transitions of the kind $\rho N \to \pi N \to \eta N$.

\begin{figure}
\begin{center}
\hspace*{-0.5cm}\epsfig{file=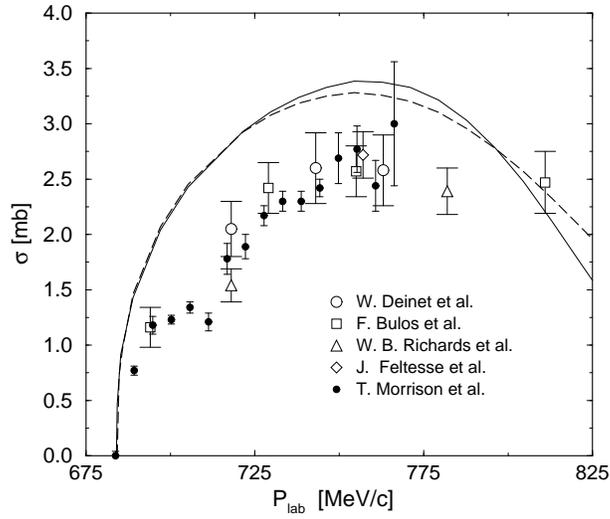, width=8.0cm} 
\end{center}
\caption{S-wave cross section for the reaction $\pi^-p\to \eta n$.
The dashed curve represents the results of 
the original $\pi N$ model of Refs.~\protect\cite{Olipap,Olithes}
whereas the solid line is the result of the extended model
that includes the additional diagrams of Fig.~\ref{rhodiag}. 
Data are taken from Ref.~\protect\cite{Baldi}. 
}
\label{pietaXs} 
\end{figure}

 
In view of the importance of the $\rho N \to \eta N$ transition
amplitude in many of the earlier investigations of $\eta$ production,
cf. the Introduction, and specifically in view of the crucial role
it plays in the analysis of F\"aldt and Wilkin \cite{Wilnew} we
want to explore how well this amplitude is constrained within our
coupled-channels model and to which extend it can be varied. 
For this purpose 
we have created a variant of the $MN$ model where we now
explicitly include a coupling of the $\rho N$ channel to the
$S_{11}$ $N^*(1535)$ resonance, cf. the diagrams in 
Fig. \ref{rhodiag}.  Evidently, we need to make sure that the
description of the $\pi N$ data in the region not far from the $\eta$ 
threshold remains basically unchanged when those diagrams are added
to the model. The bare coupling constants $g_{\pi NN^*}$, $g_{\eta NN^*}$, 
and $g_{\rho NN^*}$ have been varied to explore the variation 
in the amplitude for $\rho N \to \eta N$
while staying as close as possible to the experimental 
phase shift and inelasticity of the $S_{11}$ $\pi N$ partial wave 
(cf. Fig. \ref{S11rho}) and the $\pi N \to \eta N$ transition 
cross section produced by the original model 
(cf. Fig. \ref{pietaXs})  
in a region of about 60 MeV around the $\eta N$ threshold.

%
All other $\pi N$ partial waves 
remain unchanged since the parameters of the t- and u-channel exchange 
potentials were not varied. 

Eventually we settled on a small and negative value for the bare 
coupling constant $g_{\rho NN^*}$ since  
it turned out that only choosing it to be negative
allows to achieve a significant influence of those additional diagrams
on the $\rho N \to \eta N$ amplitude and, in turn, on the
predictions for $\eta$ production in $NN\to NN\eta$ reactions.
In Sec. \ref{Results} we present results for this extended
model together with those based on the original $MN$ model.

As is evident from Fig.~\ref{pietaXs} the present $MN$ model of
the J\"ulich group overestimates the $\pi^-p\to \eta n$ cross section
by roughly 15\%. In order to estimate the impact of this shortcoming
on the results for $NN\to NN\eta$ we performed some exploratory
calculations based on a reduced $\pi N \to \eta N$ amplitude. 
Specifically, we weakened the 
$\pi N \to \eta N$ transition potential of the $MN$ model \cite{Olipap}
phenomenologically by decreasing the coupling of the
$a_0$ meson at the $a_0\pi\eta$ vertex \cite{Sibi} 
so that the transition cross section is reproduced. (Note that  
the $\pi N$ phase shifts are no longer described in this case!)  
It turned out that a decrease in $\sigma_{\pi^-p\to \eta n}$ by
15\% implies a reduction in the predictions for $\sigma_{NN\to NN\eta}$ 
by roughly the same amount. 


\subsection{Vertex parameters}
\label{Par}

Besides the $MN \to \eta N$ amplitude the production amplitude
$A_{2\to 3}$ also contains the meson-nucleon-nucleon ($MNN$) vertices 
from where the rescattered mesons are emitted, cf. Fig. \ref{diag}a. 
In this subsection we want to
provide the parameters involved in those vertices, i.e. 
the coupling constants and the cutoff masses in the form factors.
As far as the coupling constants are concerned
all of them are taken over from the full Bonn $NN$ model \cite{MHE}. 
The only exception is the one of the $\eta$ meson -- because 
this meson is not included in the full Bonn model. 
Here we take the value which follows from SU(6) symmetry,
i.e. $g_{\eta NN}^2/(4\pi)=1.8$, which is also used in the $MN$ 
model \cite{Olipap} for the $\eta NN$ vertex in the 
nucleon-exchange diagrams. 

The vertex form factors are likewise taken over
from the full Bonn $NN$ model or (in case of the $\eta$ meson) from
the $\pi N$ model \cite{Olipap}. In those models the form 
factors are assumed to be of conventional monopole form, i.e.
$ F({\vec q}_M)=(\Lambda_{MNN}^2-m_M^2)/
(\Lambda_{MNN}^2+\vec{q}_M^{\: 2})$, 
where
$\Lambda_{MNN}$ is the cutoff parameter and $\vec{q}_M$ and $m_M$ are the 
3-momentum and mass of the exchanged meson, respectively. 
The values of the employed vertex parameters
are summarized in Table \ref{paramet}. 


Since the elementary $MN \to \eta N$ amplitude and also the 
vertex parameters are fixed from earlier investigations we
do not have any free and/or adjustable parameters in the production 
amplitude $A_{2\to 3}$. Thus, our results for the reaction
$NN \to NN\eta$ are genuine model predictions.

\subsection{$NN$ interaction in the initial state}
\label{ISIeffects}

The laboratory energy corresponding to the $\eta$-production threshold
is $T_{lab}=1250$ MeV. Thus, the $NN$ interaction in the initial
state takes place at rather high energies. At such energies the
effects of the ISI are characterized by the following
features:
\begin{itemize}
\item{The ISI has practically no influence on the energy dependence of the 
$\eta$ production cross section because 
the variation of the $NN$ interaction in the initial state within the energy 
interval determining the threshold region is negligible \cite{said}.} 
\item{The $NN$ scattering in the energy region in question is already 
strongly inelastic because multiple-pion production channels are open. 
In Ref. \cite{H-N} it is shown that in such a case the ISI leads to a
significant reduction of the cross section. 
}
\end{itemize}

\begin{figure}[h]
\begin{center}
  \vspace{0.5cm}
\hspace{-1.5cm}\epsfig{file=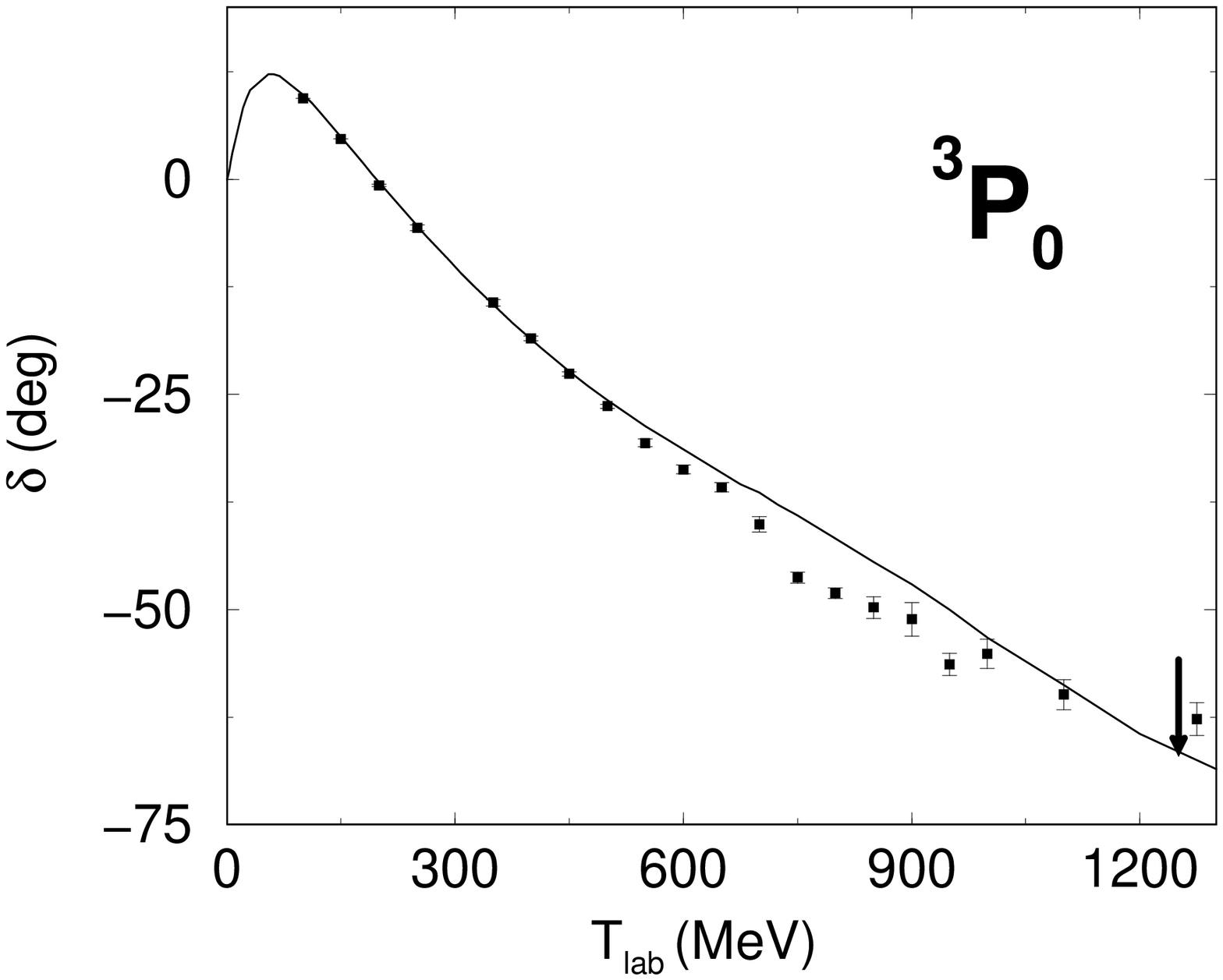,width=7.7cm}
\epsfig{file=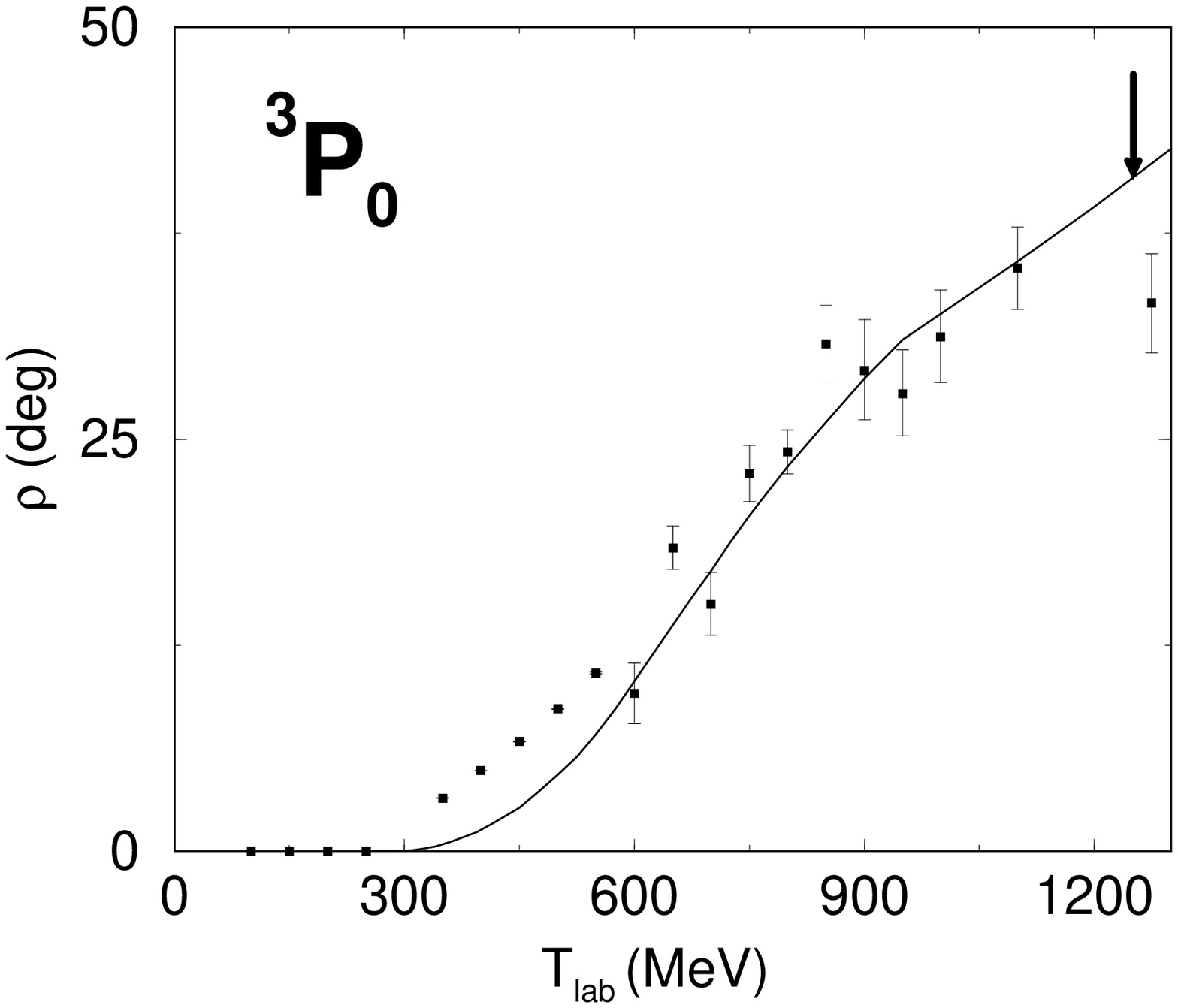,width=7.4cm}
\vspace{0.5cm}
\caption{Phase shift $\delta$ and inelasticity parameter $\rho$ 
($\eta=cos^2{\rho}$) calculated for the $^3 P_0$ partial wave 
using the CCF model \protect\cite{HaidCCF}. 
The squares represent experimental phase shifts, extracted from the SAID 
library \protect\cite{said} and recalculated from the Arndt-Roper
convention \protect\cite{AR} to satisfy the condition for the 
S-matrix: $S=\eta e^{2i\delta}$. The arrows 
indicate the $\eta$ production threshold.
}
\label{3p0}
\end{center}
\end{figure}

\begin{figure}[h]
\begin{center}
\vspace{0.5cm}
\hspace{-1.5cm}
\epsfig{file=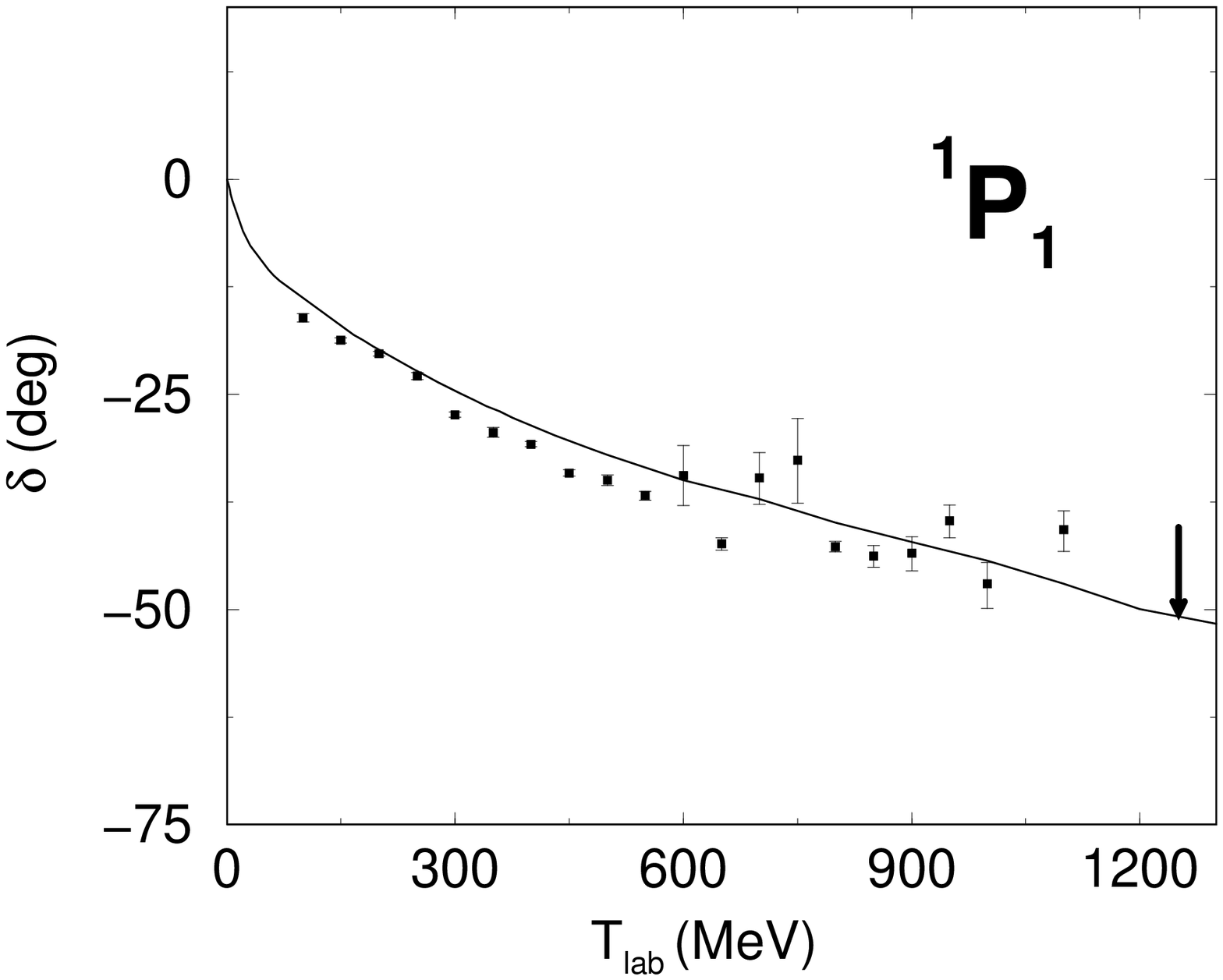,width=7.7cm}
\epsfig{file=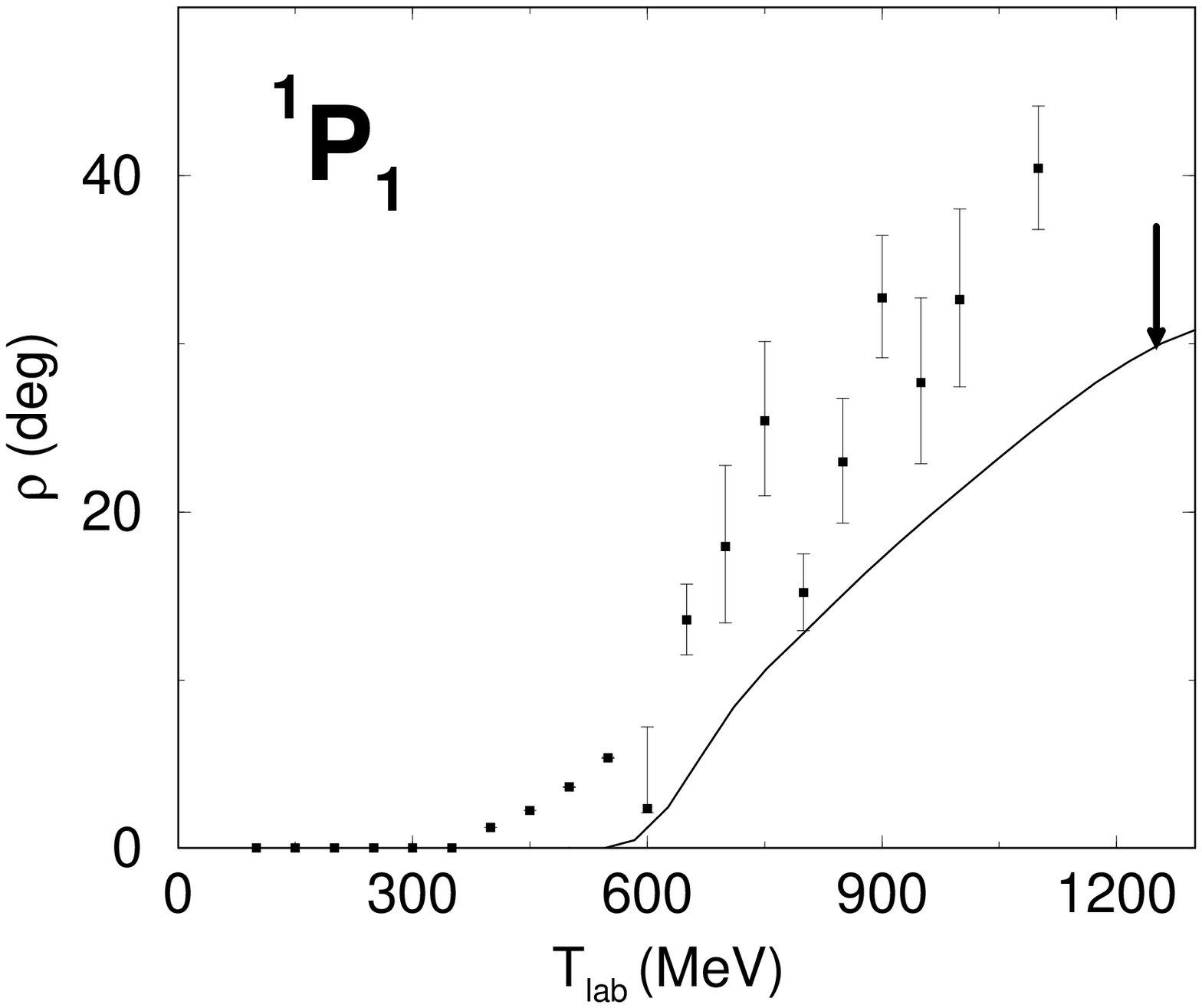,width=7.4cm}
  \vspace{0.5cm}
\caption{Phase shift $\delta$ and inelasticity parameter $\rho$ (
$\eta=cos^2{\rho}$) calculated 
for the $^1 P_1$ partial wave using the CCF model \protect\cite{HaidCCF}. 
Same description of squares as in Fig. \ref{3p0}.}
\label{1p1}
\end{center}
\end{figure}

Indeed, the energy relevant for the ISI is so high that the well known 
realistic $NN$ models such as 
the Bonn and the Paris potentials \cite{MHE,Lac} can not be applied anymore.
In case of the Paris or the Bonn one-boson-exchange (OBE) potentials 
no inelastic 
channels were included and therefore they are not valid for energies 
above $\pi$ production threshold. 
In the full Bonn model such inelastic channels are, in principle, 
build in via couplings to the $N\Delta$ and $\Delta\Delta$ channels. 
However, since meson retardation 
effects are retained as well three body singularities occur 
for energies above $\pi$ production
threshold and hence the calculation is technically much more 
difficult \cite{Elst2}.
Therefore, in the following we use an alternative 
$NN$ model, CCF \cite{HaidCCF}, which contains the same dynamics as the 
full Bonn model, specifically the coupling to 
$N\Delta$ and $\Delta\Delta$ channels, but where meson retardation effects are 
removed by the so-called folded diagrams technique.
In Figs. \ref{3p0} and \ref{1p1} we present the results of the CCF 
model \cite{HaidCCF} for the $NN$ phase shifts and 
inelasticities in the $^3\! P_0$ and $^1\! P_1$  partial waves. Since
we restrict ourselves to final states composed purely out of $s$--waves, these
are the relevant partial waves for the $NN$ interaction in the inital state. 
The agreement with the experimental data is quite good, specifically in view
of the fact that the model was only fitted to the $NN$ phase shifts below
the pion production threshold, i.e. for $T_{lab} \le$ 300 MeV. 
Note, that
in this model the inelasticities are introduced only through the 
coupling to channels involving the $\Delta$ isobar. 
Therefore, in the $^1\! P_1$ channel, where only the simultaneous 
exitation of two $\Delta$ isobars is allowed, the inelasticity becomes noticable
only after 600 MeV kinetic energy, i.e. somewhat later than indicated by
the empirical data. However, in the energy range
relevant for $\eta$ production the description is reasonable.
Finally, let us mention that the width of the $\Delta$ isobar is taken into
account phenomenologically by using a complex $\Delta$ mass in the propagator. 
Specifically, we employ a parametrization of the width given by Kloet and
Tjon in Ref.~\cite{Kloet} which is energy- as well as momentum-dependent. 


The evaluation of the ISI effects amounts to performing a loop integration,
cf. Eq.~(\ref{M23}) and Fig.~\ref{Mamp}, which gives rise to
contributions from the unitarity cut and from the principal value (PV) integral.
Since in the initial state the colliding nucleons 
have rather large relative momenta 
it was argued in Ref.~\cite{H-N} that the contribution of the PV integral 
should be suppressed as compared to the piece resulting from the on-shell 
(unitarity) cut. 
Under this assumption the ISI effect on the cross section reduces to a 
simple multiplicative factor $|\lambda|^2$ which can be expressed in terms of 
on-shell $NN$ information only, namely phase shifts ($\delta$) and 
inelasticities ($\eta$): 
\be
|\lambda|^2=\bigl |1-i\frac{\pi E_{p}}{2}p T_{NN}\bigr |^2=
\frac{1}{4}\bigl |1+\eta \, cos(2\delta)+i\eta \,
 sin(2\delta)\bigr |^2.
\label{isi2}
\ee
Here, the T-matrix is defined as $T=-\dis\frac{2}{\pi E_p} 
\dis\frac{\eta \, e^{2i\delta}-1}{2ip}$ and $p$ ($E_{p}$)
is the cms momentum (energy) of the nucleons.

In our actual calculations it turns out that there is a
strong cancellation between the Born term and the piece coming from the
unitarity cut, i.e. between the terms corresponding 
to the ``$1$'' and to ``$\eta \, cos(2\delta)$'' in Eq.~(\ref{isi2}). 
Then the ISI factor $\lambda$ in Eq.~(\ref{isi2}) is rather small. But in such
a case, the contribution from the PV integral, which is certainly small and 
which was neglected in Eq.~(\ref{isi2}), becomes important as well and has an
influence on the ISI effects. We will come back to this point in the
next section. 

Due to 3-body singularities appearing in the loop with the ISI the 
evaluation of the PV integral is very involved~\cite{Motzke}. 
In the present work we avoid the technically rather tedious 
explicit treatment of this 3-body singularity.
Rather we follow the approach which was already used 
in the work of Batini\'c  et al. \cite{Bat} and suppress the 
3-body singularity by putting the nucleon energies appearing in
the meson propagator in the intermediate state on the energy shell. 
We want to emphasize that this approximation still provides the correct value 
for the contribution from the two body $NN$ unitarity cut in the ISI loop.
Only the contribution from the PV integral is influenced by 
the approximation discussed above.

The effects of the ISI on the results for $\eta$ production in 
$NN$ collisions will be discussed in detail in section
\ref{Results}. Here we want to emphasize only that  
the PV integral in the ISI loop introduces an interesting feature. 
It breaks the universality of the ISI as it is suggested by the 
prescription given in Eq.~(\ref{isi2}), i.e. it implies that
the reduction caused by the ISI will differ for the various 
contributions to $\eta$-meson production and, thus, depends also on the 
production mechanism. The approximative treatment of ISI
according to Eq.~(\ref{isi2}) is unable to account for such effects.

\subsection{$NN$ interaction in the final state}
\label{FSI}

Effects of the $NN$ interaction in the final state in various
production reactions were investigated in detail in 
Refs. \cite{H-N,BarFSI,Wat,Mig,Nis}. 
In particular, in Ref. \cite{BarFSI}
it was demonstrated that the $NN$ FSI cannot be factorized from the 
production amplitude if one wants to obtain
reliable quantitative predictions for the cross sections. 
This conclusion confirms the arguments given 
in Ref. \cite{H-N}. Also it was shown in Ref. \cite{BarFSI} 
that the use of the Jost function of some 
realistic $NN$ potentials for the evaluation of FSI effects is 
invalid and may lead to results 
considerably different from the ones based on a proper calculation. 
In this work $NN$ FSI effects are treated consistently, i.e.
the calculation of loop integrals is performed with taking 
into account the full dynamics both from the $NN$ T-matrix
and from the production amplitude (cf. Eq.~(\ref{M23}) or Fig.~(\ref{Mamp}),
respectively). 
In the calculation we utilize not only the CCF \cite{HaidCCF} $NN$ model, 
i.e. the one which is used for the ISI as well, but also 
Bonn B \cite{OBEPQB} in order to examine
the sensitivity of the results to the $NN$ interaction model. 
Both models describe equally well the $NN$ phase shifts in the 
$^1 S_0$ and $^3 S_1$ partial waves, which are relevant for the $\eta$ 
production, for energies below the pion production threshold, i.e. 
in the energy region relevant for the FSI.

Note that the Coulomb interaction in the $pp$ final state is taking into 
account, applying the prescription described in Ref.~\cite{Han1}.

\section{Results of the calculation of the reaction $NN\to NN\eta$}
\label{Results} 

\begin{figure}
\begin{center}
\vspace*{-0.6cm}
\hspace*{-0.2cm}
\epsfig{file=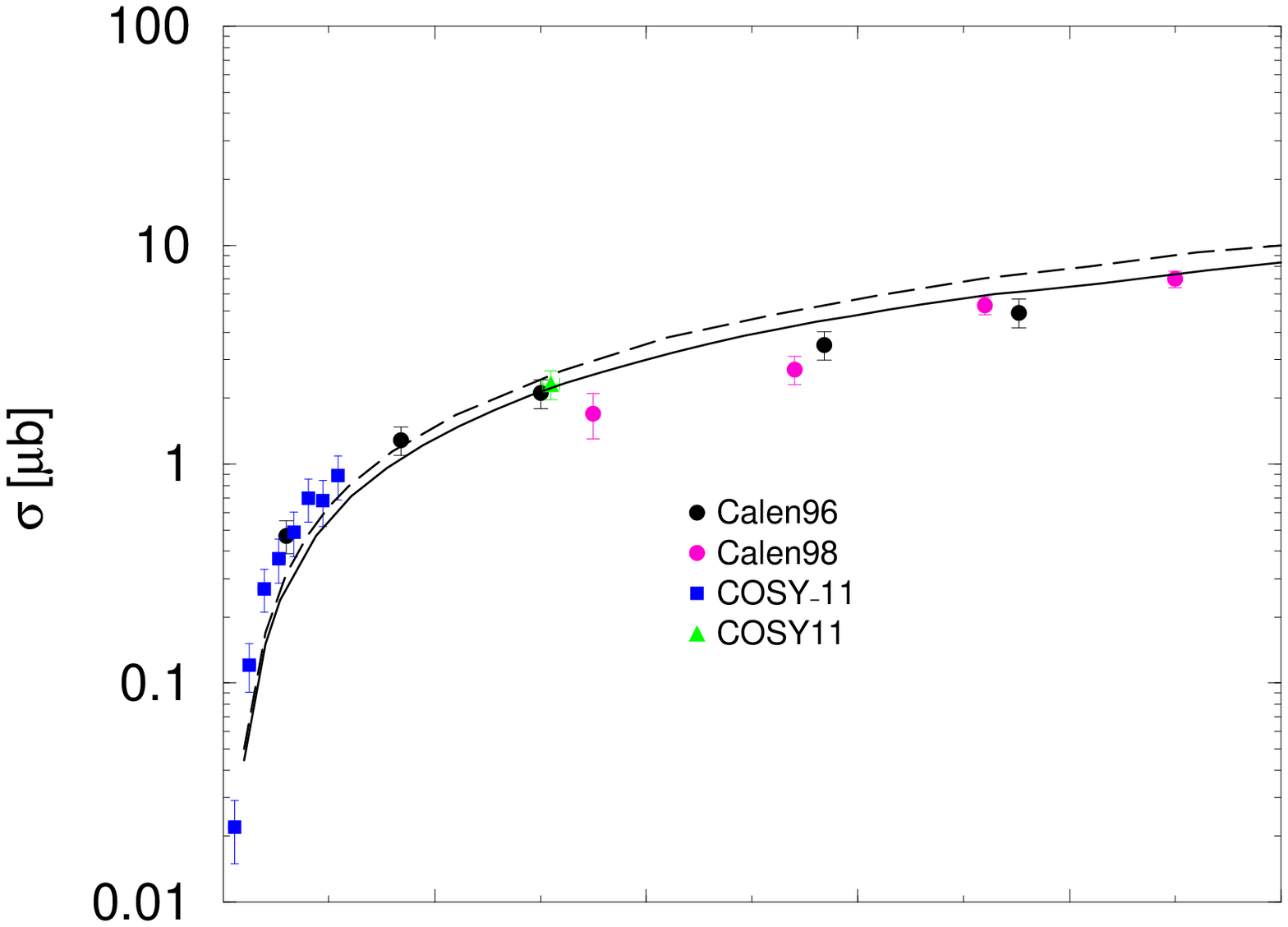, width=9.3cm}
\hspace*{0.1cm}\epsfig{file=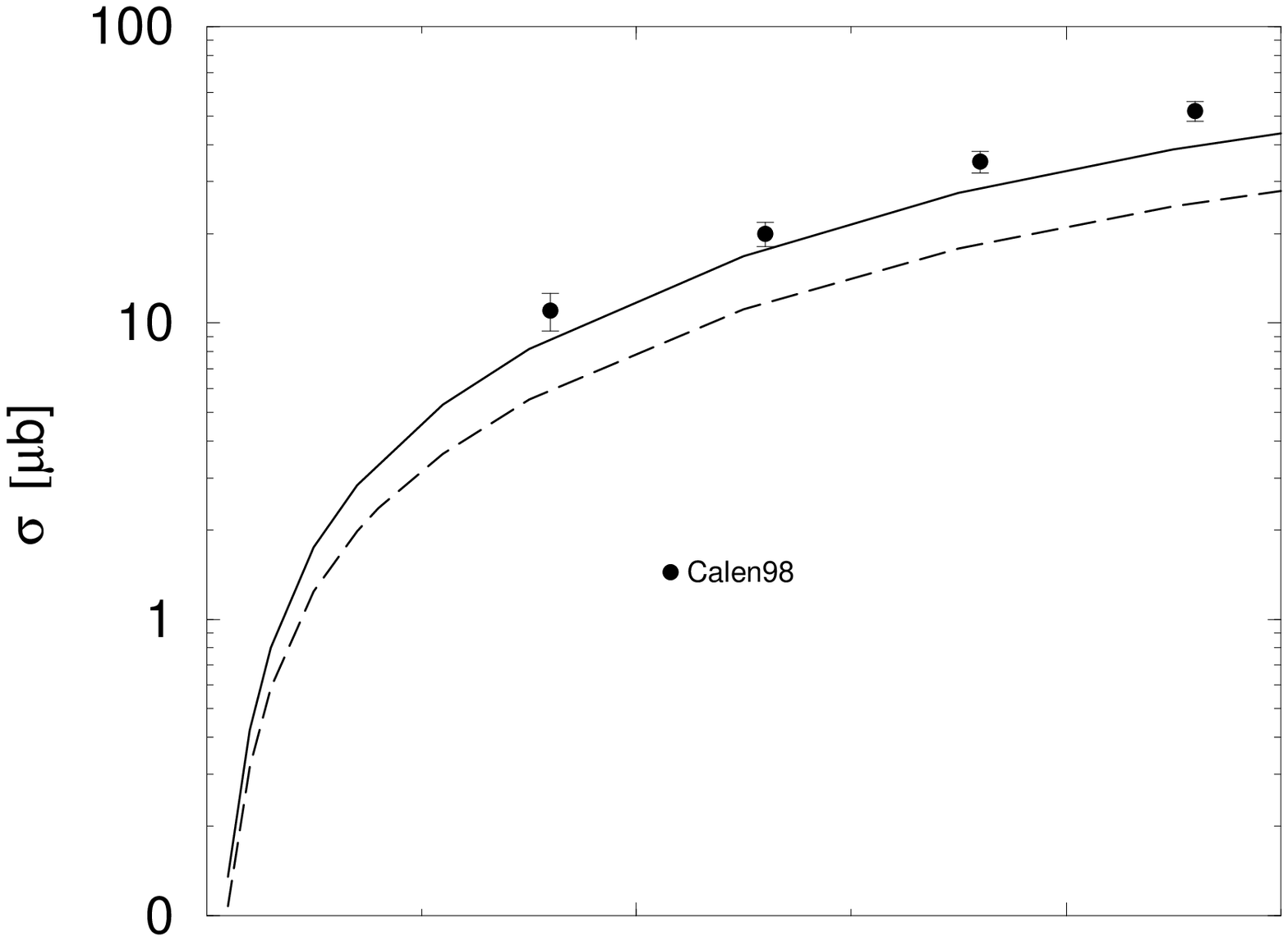, width=9.2cm} 
\hspace*{0.1cm} \epsfig{file=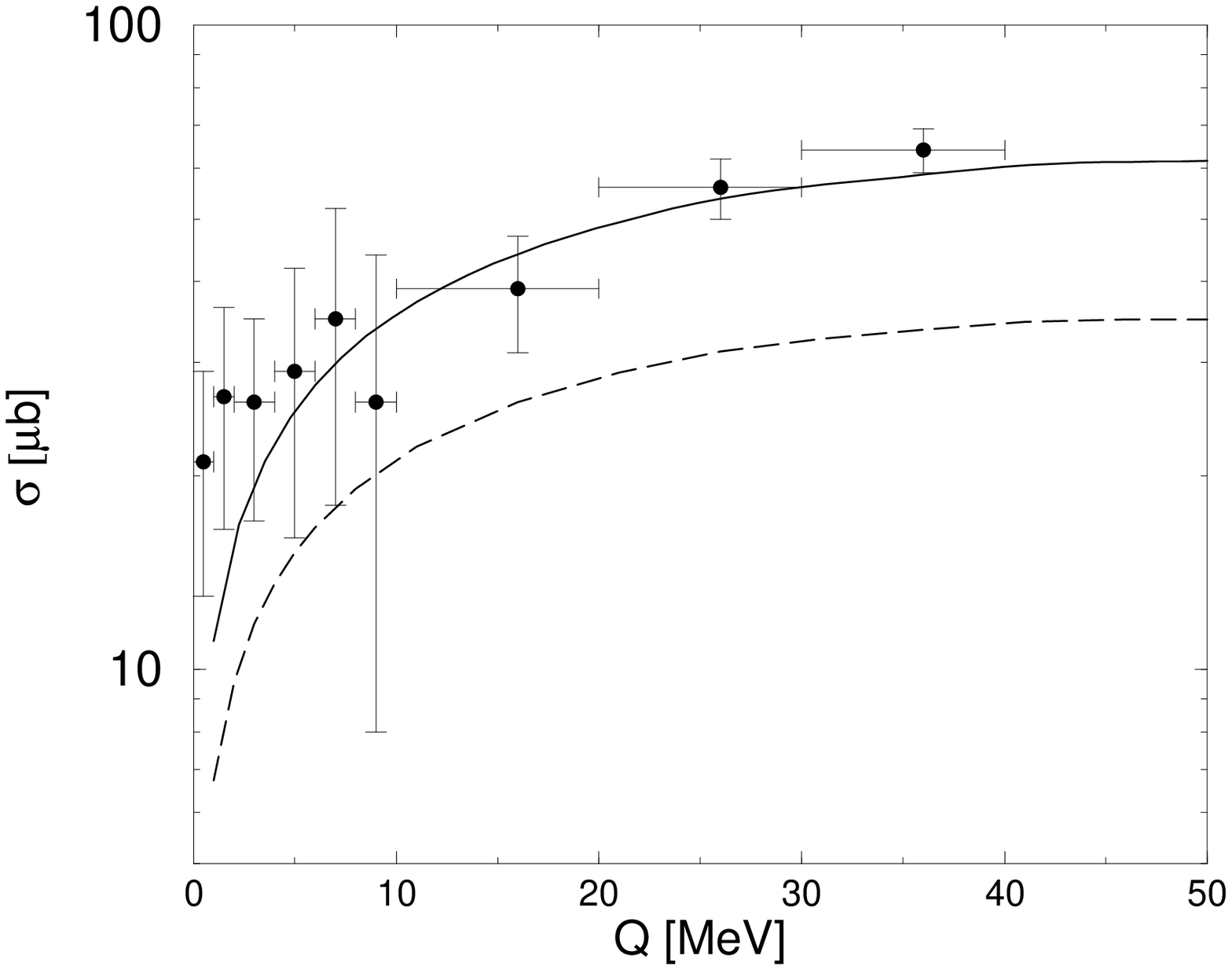, width=9.3cm}
\end{center}
\caption{Total cross sections of the reactions $pp\to pp\eta$ (top), $pn\to pn\eta$ 
(middle) and $pn\to d\eta$ (bottom).
The dashed lines represent the results of our calculation
with the original $MN$ model of Ref. \protect\cite{Olipap}, whereas the solid 
lines are based on the extended $MN$ model described and discussed
in Sect. II A. Data are taken from 
Refs. \protect\cite{Ups,COSY,COSY1,Calenpneta,Upspndeta1,Upspndeta2}.}
\label{fullxs}
\end{figure}

Our results for the reactions $pp\to pp\eta$, $pn\to pn\eta$ and $pn\to d\eta$
are presented in Fig. \ref{fullxs} and compared with empirical information
from the Uppsala/Celsius \cite{Ups,Calenpneta,Upspndeta1,Upspndeta2}
and J\"ulich/COSY \cite{COSY,COSY1} accelerator facilities.
The dashed curves correspond to the 
calculation based on the original $MN$ model 
(cf. section \ref{MN}) for the elementary $\eta$-production 
amplitude and the CCF $NN$ model \cite{HaidCCF} for the ISI and FSI.

Evidently, the calculation based on the original J\"ulich $MN$ 
model \cite{Olipap}
yields a qualitative overall description of the experimental data. 
This has to be certainly considered as success because, as mentioned before,
there are no adjustible parameters in our calculation of $NN\to NN\eta$.
As one can observe from Fig.~\ref{fullxs}, for the 
$pp\eta$ case we overestimate the cross section by approximately 30\% 
whereas for $pn\to pn\eta$ the model calculation underestimates 
the experimental data by about 50\%. 
The situation for the reaction $pn\to d\eta$ is very similar to $pn\to pn\eta$ 
since both processes are governed by the $I=0$ isospin channel. 
As is known from earlier investigations \cite{Wilnew,Nakayama}
for the latter reaction the contribution of the 
$I=1$ channel is much smaller than the one for $I=0$.


However, it is interesting to investigate the sensitivity of our results 
to the $MN$ amplitude and, specifically, to see whether the model predictions
for $NN\to NN\eta$ can be improved by modifications in the $MN$ model 
via introducing some additional graphs, as described in section \ref{MN}.
For the original $MN$ model the amplitudes 
of the $\pi$- and $\rho$-exchange contributions to $\eta$ production
turned out to be almost orthogonal to each other. 
Thus, there is practically no interference between these contributions. 
The additional
diagrams which we introduced into the $MN$ model allow a slight modification of 
the orientation of these amplitudes in the complex plane and, as a consequence,
now interference effects do occur. Specifically, it is possible to generate a 
destructive interference in the isotriplet channel ($I=1$) and a constructive
interference in the $I=0$ case. This leads to a slight reduction of the
cross section in the reaction $pp \to pp\eta$ and to a significant
enhancement for $pn \to pn\eta$, cf. the solid lines in 
Figs. \ref{fullxs}, bringing now the results quite close to the experiment. 
Also the prediction for the reaction $pn \to d\eta$ is now in 
good agreement with the experimental data. 

\subsection{Contributions of individual meson exchanges.}
\label{fullcon}

Let us examine our model in more detail and analyze the contributions of the
individual exchange diagrams. 
We do this for the calculation based on the extended $MN$ model. 
The main differences between the results obtained for 
the original $MN$ model and its extended version 
will be discussed at the end of this section. 

\begin{figure}
\begin{center}
\epsfig{file=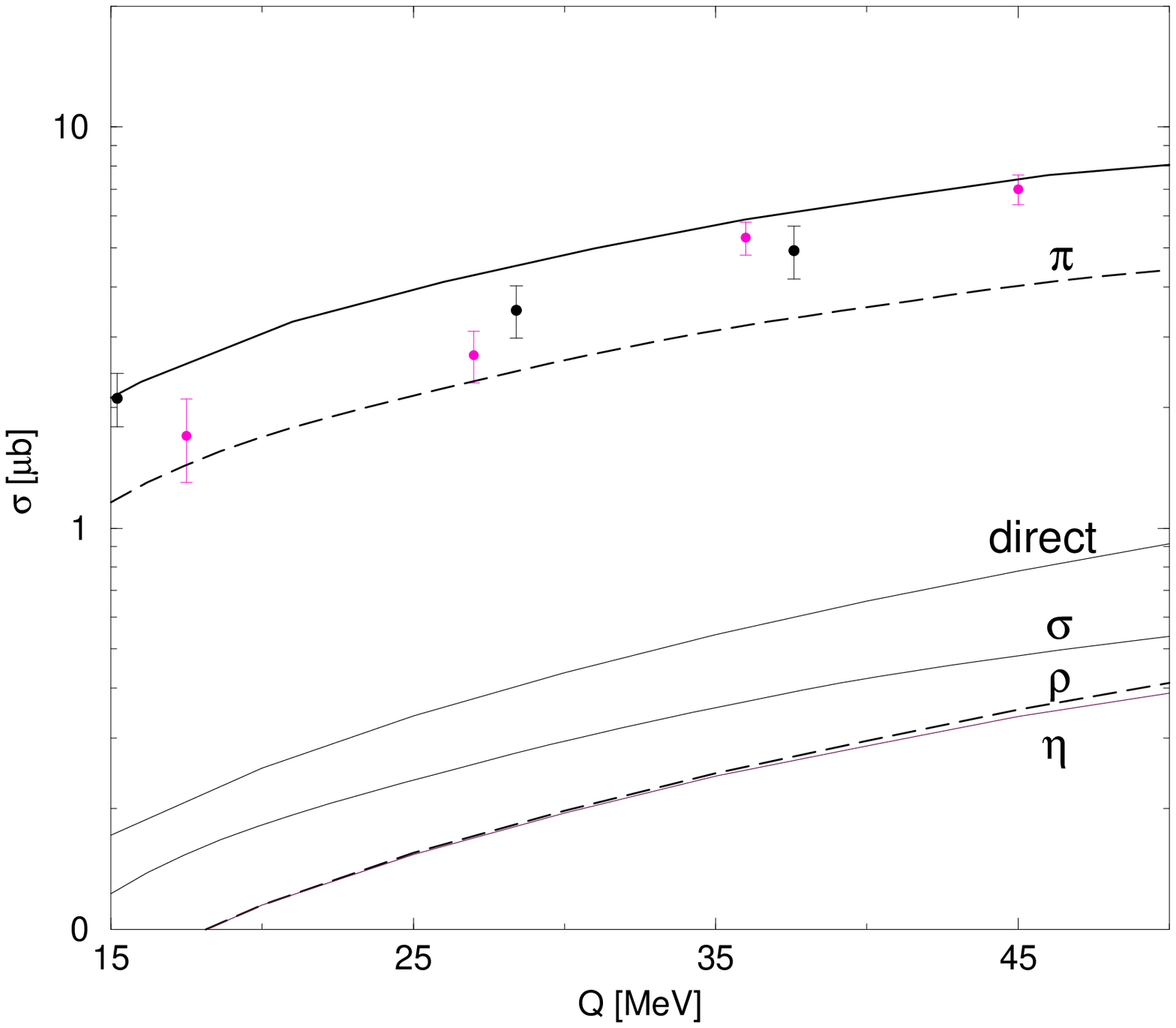, width=11.0cm} 
\end{center}
\caption{Contributions of the individual meson exchanges to the 
reaction $pp\to pp\eta$.
The presented results are based on the extended $MN$ model.
The solid line is our full result, i.e. when all contributions are summed
up coherently.}
\label{fullcontrppeta}
\end{figure}

\begin{figure}[h]
\begin{center}
  \hspace{-0.2cm} \epsfig{file=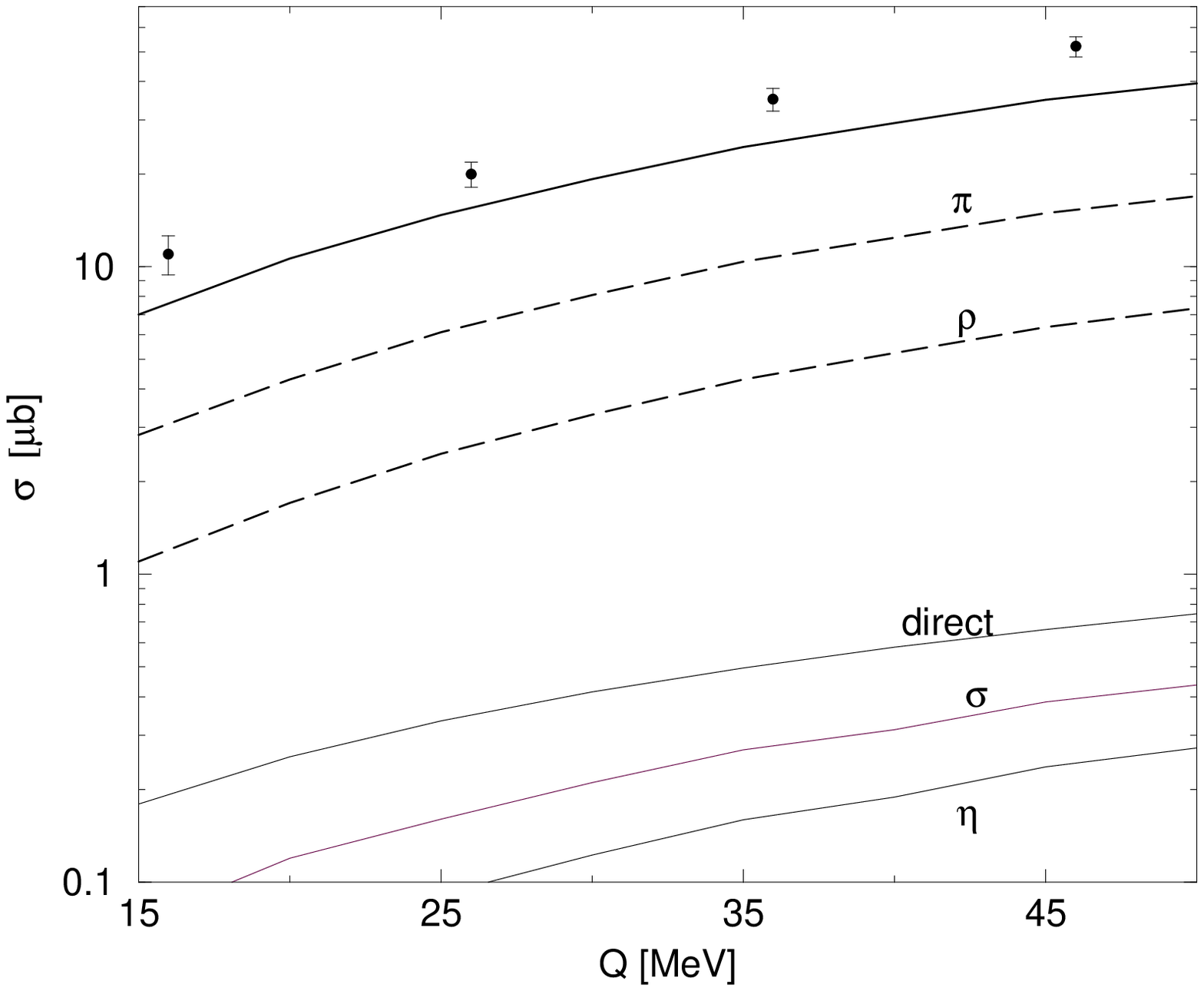, width=11.2cm} 
\end{center}
\caption{Contributions of the individual meson exchanges to the 
reaction $pn\to pn\eta$ ($I=0$). 
The presented results are based on the extended $MN$ model. 
The solid line is our full result, i.e. when all contributions are summed
up coherently.}
\label{fullcontrpneta}
\end{figure}

Corresponding results are presented in 
Fig.~\ref{fullcontrppeta} ($pp\to pp\eta$) 
and in Fig.~\ref{fullcontrpneta} ($pn\to pn\eta$ for $I=0$),
where we show the production cross section for the individual
meson exchanges together with the full results. 
Furthermore, in Table \ref{contribISI} we compare cross sections 
of the full calculation with results obtained without ISI,
i.e. where $T_{NN}^{ISI}$ in Eq.~(\ref{M23}) was set to zero, 
at the specific excess energy $Q$ = 35 MeV. 
The latter allows us not only to expose the relevance of the various meson 
exchange (rescattering) contributions for the $\eta$ production cross
section but also to elucidate that their relative importance is 
strongly influenced by the ISI. For example, in the model calculation
without ISI the dominant contributions come from $\pi$ and $\rho$
exchanges, both being of comparable magnitude. However, the ISI reduces
the $\rho$-exchange contribution much more strongly then the one from
$\pi$ exchange (cf. Table.~\ref{contribISI}) leaving $\pi$ exchange
as the only dominant production mechanism. The contributions from
the other production mechanisms ($\sigma$ exchange, $\eta$ exchange, etc.)
are already comparably small before including the ISI and they are
also significantly reduced (though not as much as the $\rho$) 
by introducing the ISI. 
%

Let us emphasize in this context that phenomenological treatments like
those based on Eq.~(\ref{isi2}) are unable to account for such effects 
resulting from details of the dynamics. 
They lead only to an overall reduction of the cross 
section independent of the production mechanisms, as is indicated by the last
3 columns of Table~\ref{contribISI}. (Note that the reduction factors
used in Ref.~\cite{Nakayama} differ from those for CCF because the ones
in the former work are based on the $NN$ phase shifts listed at 
the SAID library \cite{said}. The reduction factors of 
Ref.~\cite{Wilnew} are not obtained from Eq. (\ref{isi2}) but 
from a different prescription.)

We also want to mention that the ISI effects seen in our investigation 
are at variance with those reported in the only other study of
$\eta$-meson production where the ISI was taken into account
explicitly \cite{Bat}. In that work by Batini\'c et al. it was
found that the combined distortions from the FSI and ISI are not 
sensitive to the dynamics of the production operator.
We don't have an explanation readily at hand for this discrepancy.
We can only conjecture that  
it might be due to differences in the employed  
$MN \to \eta N$ transition amplitudes.

As we have discussed above, after inclusion of the ISI 
$\pi$ exchange plays the dominant role in our 
calculation of the reaction $pp\to pp\eta$ (see Figure \ref{fullcontrppeta}). 
Nevertheless, the contributions of the other mesons, 
especially of the $\eta$ and $\rho$ exchanges 
and of the direct term are still significant due to interference effects 
with the amplitude of the $\pi$ exchange.
The $\eta$ exchange and the direct term contribute constructively to the
reaction $pp\to pp\eta$ whereas the interferences from the 
$\sigma$ and $\rho$ mesons are destructive. 

Note that the actual effect of those interferences depends also on 
whether the ISI is included or not. That can be best seen by the 
fact that the ISI reduces the individual meson-exchange contributions by
factor 4 or more whereas the total cross section is only reduced by 
roughly a factor 2, cf. Table~\ref{contribISI}. 

Let us now come to the reaction $pn\to pn\eta$.
The cross section of this reaction is primarily determined by the 
isovector particles ($\pi$ and $\rho$) in the $I=0$ channel,
because their contributions are weighted by the 
large isospin coefficient ($9$ versus $1$ for $I=1$).
In addition, for the relevant $NN$ partial wave in the initial state, 
the $^1P_1$, it turned out that the variation of ISI effects for the 
various  production mechanisms is not as pronounced as in 
the $I=1$ ($^3P_0$) case. Indeed, except for pion exchange, 
in this channel the reduction due to the ISI is almost universal 
as can be seen in Table~\ref{contribISI}. 


The relative contributions of the individual meson exchanges 
to the reaction $pn\to pn\eta$ (in the $I=0$ channel) are 
shown in Fig. \ref{fullcontrpneta} for the full calculation.  
As mentioned already,
the dominant role belongs again to the pion exchange, though the contribution
from $\rho$ exchange is now much less suppressed as in the $I=1$ case (cf. Fig.
\ref{fullcontrppeta}). The individual contributions from the other production 
mechanisms are again much smaller. 
Like in case of $pp\to pp\eta$ interferences
play a role once we add all contributions coherently. 
Specifically, the
$\eta$- and the $\rho$-meson exchange exhibit opposite features as in the 
$pp\to pp\eta$ case, i.e. the $\eta$ is destructive while the $\rho$ is constructive. 
The direct term acts again constructively and its contribution nearly cancels 
with the one resulting from $\eta$ exchange.
The contribution from $\sigma$ exchange turns out to be negligible. 
  
The interference effects between $\pi$ and $\rho$ meson exchange amplitudes 
for the original and extended $MN$ models are illustrated in 
Table \ref{pirhocontr}, where we present the ratio of the coherent
sum of the two amplitudes to the incoherent sum. For the original 
$MN$ model this ratio is close to 1 (for both isospin channels) 
verifying that there are only small interference effects. For the 
extended $MN$ model those interferences are much more pronounced. 


Table~\ref{contribISI} lists also the phenomenological reduction factor that 
follows from Eq.~(\ref{isi2}) for the employed $NN$ initial state interaction
(in the column with the header 'CCF'). Thus, we can compare directly the
reduction of the cross section that follows from the explicit inclusion of
the ISI with
the one suggested by the phenomenological prescription. It is evident that
the prescription Eq.~(\ref{isi2}) doesn't work that well for the $\pi$ and
$\rho$ exchange contributions. For the $\eta$ and specifically for the $\sigma$ 
exchange, however, the results are fairly similar. 

\subsection{Role of the $\eta NN$ coupling constant}

Now we would like to comment on the influence of the $\eta NN$ coupling 
constant. As mentioned before, in our calculation we take the value 
$g_{\eta NN}^2/(4\pi)=1.8$ that follows from SU(6) symmetry.
This value is already much 
smaller than those used in OBE versions of the Bonn $NN$
model \cite{MHE,OBEPQB}, say, which are around 3 to 7. On the other 
hand, some studies of $\eta$ production in $NN$ collisions
suggest that only still smaller $\eta NN$ coupling constants allow to 
describe the experimental data \cite{Wilnew,Pena}. E.g., a
value of only about $g_{\eta NN}^2/(4\pi)=0.4$ was employed 
in \cite{Pena} while in \cite{Wilnew} the $\eta$ contribution was even 
completely neglected. Therefore, it is interesting to see whether our
results on $\eta$ production based on the original $MN$ model could
be also improved by simply varying the $\eta NN$ coupling constant
in the production operator. 

In our model calculations $g_{\eta NN}$ enters via the $\eta$ exchange
contribution but also in the direct term. The latter is
not negligible and interferes constructively with the dominant $\pi$ exchange
contribution in both isospin channels of the reaction $NN\to NN\eta$.
On the other hand, as mentioned before, the $\eta$ exchange
contribution interferes constructively for the reaction
$pp\to pp\eta$ but destructively for the $I=0$ part of $pn\to pn\eta$.
Indeed, the direct term and the $\eta$ exchange basically cancel each
other in the dominant isospin $I=0$ channel of the reaction
$pn\to pn\eta$ and, therefore, the $\eta NN$ coupling
constant influences primarily the cross section of the $pp\to pp\eta$
reaction.
Thus a reduction of $g_{\eta NN}$ would only decrease the cross section
for $pp\to pp\eta$. It would not alter the results for $pn\to pn\eta$
and $pn\to d\eta$ and, therefore, does not lead to an improvement for the
latter two reaction channels. 

\subsection{Sensitivity to the $NN$ interaction.}

\begin{figure}
\begin{center}
\hspace*{-0.2cm}
\epsfig{file=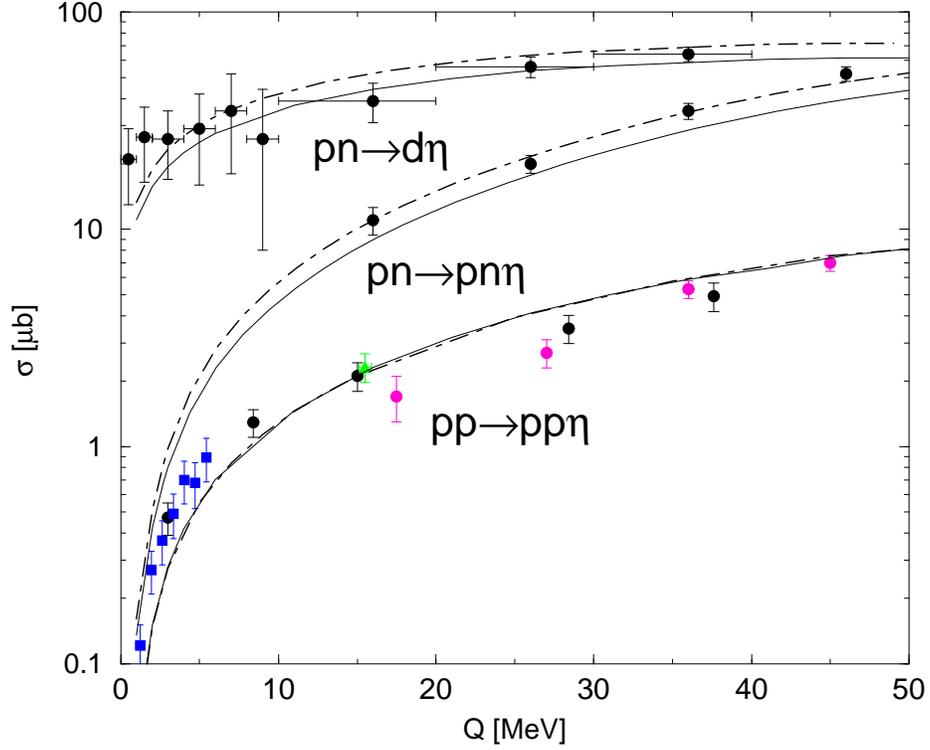, width=12.3cm}
\end{center}
\caption{Total cross sections of the reactions $pp\to pp\eta$, $pn\to pn\eta$,
and $pn\to d\eta$
employing different $NN$ models for the final state interaction.
The solid lines represent the results with the CCF $NN$ model 
\protect\cite{HaidCCF} whereas the dashed-dotted lines were obtained 
for the Bonn B model \protect\cite{OBEPQB}.
The calculations are based on the extended $MN$ model.
}
\label{CCFBonnB}
\end{figure}

In order to examine the sensitivity of our results to differences in the 
$NN$ interaction in the final state we also performed a calculation where the 
OBE potential Bonn B \cite{OBEPQB} was utilized. 
Corresponding results are presented in Fig. \ref{CCFBonnB} 
by dashed-dotted lines. The results with 
the CCF $NN$ model are also shown (solid lines) in order to facilitate 
a comparison.
As one can see from this figure, for the $pp\to pp\eta$ channel the
predictions of both $NN$ models lie basically on top of each other. 
However, for the $pn\to pn\eta$ channel the results based on the 
Bonn B potential are about 20\% larger and, indeed, are practically 
in agreement with the experiment. At the same time the cross section for
the $pn\to d\eta$ channel is also enhanced by about 20\% and now
slightly overshoots the data.  

In any case, the differences between the results for the two considered
$NN$ models are not that large. But let us emphasize here an interesting
by-product of the above comparison. Obviously it is not possible to
achieve a simultaneous description of all three measured $\eta$ production
channels. Either we have agreement for the reactions 
$pp\to pp\eta$ and $pn\to d\eta$ (for CCF) and then $pn\to pn\eta$ is off,
or $pp\to pp\eta$ and $pn\to pn\eta$ are reproduced (Bonn B) and then
the results for $pn\to d\eta$ deviate from the data. Let us remind the
reader that, close to threshold, there are only two independent amplitudes
that determine those three reaction channels. 
The same incompatibility is also seen in the results of 
F\"aldt and Wilkin, cf. Fig. 5 of Ref.~\cite{Wilnew},
and, therefore, we believe that it is not due to the specific production
mechanism employed in our study. 
It remains to be seen whether this discrepancy will disappear when 
contributions from higher partial waves are included. In any case, we
want to mention that a similar incompatibility exists also for
near-threshold pion production, cf. the results in Fig. 3 of 
Ref.~\cite{Han2} and also in Ref.~\cite{GFCW}.


\subsection{Comparison with other model calculations.}

In this subsection we want to discuss qualitatively the
differences between our model and other model calculations in the
literature. We restrict the comparison to such calculations
where also $pp$ as well as $pn$ induced $\eta$ production channels
are considered, i.e. to the works of Gedalin et al. \cite{Ged} and 
Nakayama et al. \cite{Nakayama}, where results for 
the reactions $pp\to pp\eta$ and $pn\to pn\eta$ are presented,
and to Ref. \cite{Wilnew}, where in addition the channel
$pn\to d\eta$ is considered.  

The relatively large value of 6.5 for the cross section ratio
$\sigma_{pn\to pn\eta}$/$\sigma_{pp\to pp\eta}$ established in
the CELSIUS experiment \cite{Calenpneta} revealed that the
dominant $\eta$-production mechanism has to be of isovector 
nature. (Exchange of scalar mesons would yield a ratio of
roughly one!) Furthermore, simple estimations \cite{Calenpneta,Wilnew}, 
taking into account the different strengths in the final ($pp$ versus $pn$) 
state interactions, strongly suggest that a production 
operator involving only a single isovector-meson exchange 
($\pi$ or $\rho$) might still fall short of describing the 
experiments and, thus, further production mechanisms should be 
relevant. 
Indeed, a scenario with a dominant contribution from isovector meson 
exchange is realized in all the models cited above, including ours. 
However, while $\rho$ exchange is the
dominant $\eta$ production mechanism in Refs.~\cite{Ged,Wilnew}
and in the alternative model of Ref.~\cite{Nakayama} it is
$\pi$ exchange that provides the bulk of the cross section in our
model and in the regular model of Nakayama et al.~\cite{Nakayama}.  
We should mention, however, that in our model $\pi$ and $\rho$
exchange are of pretty much the same strength and provide comparable
cross sections before the ISI is included, as can be seen from 
Table~\ref{contribISI}. If we would ignore ISI effects altogether,
as done in Ref.~\cite{Ged}, or use the phenomological prescription
employed in Refs.~\cite{Wilnew,Nakayama} this feature would persist.

The models differ also significantly when it comes to the role played
by the smaller contributions. For example in the model of 
F\"aldt and Wilkin the $\pi$ exchange is very important. Its strong
interference with the dominant $\rho$ exchange contribution is
the main mechanism which allows those authors to achieve 
a consistent description of the experimentally observed 
cross sections for the reactions $pp\to pp\eta$ and ${pn\to pn\eta}$.
On the other hand, in the alternative model of Nakayama \cite{Nakayama}
it is the interference of the $\omega$ exchange with the dominant 
$\rho$ exchange that is responsible for 
obtaining a reproduction of the data. Finally, in the regular
model of the latter publication agreement with the data is achieved 
by the interference of the $\eta$ exchange with the dominant
$\pi$ exchange.
 
In our model calculation based on the extended $MN$ model it is 
the constructive interference between the $\pi$- and $\rho$-meson 
contributions which yields the main enhancement in the 
cross sections for $pn\to pn\eta$ and $pn \to d\eta$
and brings the results close to the experiment. 
Thus, for the 
$I=0$ dominated reactions the mechanism is similar to the one 
in the model of F\"aldt and Wilkin. However, this is no longer
true for the reaction $pp\to pp\eta$. Here, in our model the 
$\pi$--$\rho$ interference does not play an important role. 
The reason for that are the strong effects from the ISI, 
discussed above, which reduce the $\rho$-exchange contributions
much more strongly than those coming from $\pi$ exchange, cf.
Table~\ref{contribISI} and, therefore, suppress also the 
interference. The reproduction of the data for the 
reaction $pp\to pp\eta$ is mainly due to a constructive
interference of the dominant $\pi$-exchange contribution with
the contributions from $\eta$ exchange and from direct $\eta$
production. Thus, for the $I=1$ case the mechanism is similar to the one 
in the regular model of Nakayama \cite{Nakayama}. 

In the work of Gedalin et al.~\cite{Ged} several interference patterns
are studied. But, in general, their results do not  
agree that well with the experiments in the energy region 
$Q \le 50$ MeV considered. Because of those reasons we refrain from
a more detailed comparision with their results. Let us mention, however, 
that the most prominent feature in their model is the absence of any
interferences between the contributions of vector meson exchange 
and pseudoscalar (or scalar) meson exchange. The orthogonality of 
those contributions is achieved by a peculiar choice of interaction
Lagrangians for the vector mesons. However, it seems to us that those
Lagrangians do not fulfil the requirement of time reversal invariance.  
Also, they seem to be in contradiction to the ones applied by these 
authors in their own earlier investigations \cite{Moal}.

Interestingly, also in our calculation based on the original $MN$
model there is practically no interference between the $\pi$ and
$\rho$ exchange contributions, as already mentioned above. 
But in our model the origin of this feature is quite different. 
Here the orthogonality of the two amplitudes is generated by the
dynamics of the coupled-channels $\pi N$ model. In particular, the
Born term $\rho N \to N^*(1535) \to \eta N$ that is the 
main source of the orthogonality in the model of Gedalin et al. 
is not even present in the original $MN$ that we use in our study
of $\eta$ meson production. Once this diagram is included 
in the coupled-channels $MN$ model 
(cf. the extended $MN$ model described in subsect.~\ref{MN}) 
interference effects become more pronounced rather than suppressed.

\section{Summary}

We performed a detailed theoretical calculation of the different 
channels of the reaction $NN\to NN\eta$  
($pp \to pp \eta$, $pn \to pn \eta$, $pn \to d\eta$) in the near-threshold 
region, i.e. for excess energies up to about 50 MeV.
The production mechanisms which have been included
consist of re-scattering terms with $M=\pi,\rho,\eta,\sigma$ meson exchanges 
and the direct $\eta$ production.
Effects of the final and initial state interaction between the nucleons are 
fully taken into account. The calculation utilizes the CCF $NN$ model for the 
treatment of the FSI and ISI and 
a realistic coupled-channels model of the $\pi N$ system 
for the evaluation of the $MN\to \eta N$ transition amplitudes. 
A qualitative agreement of calculated cross sections with the 
experimental data is achieved for all considered $\eta$ production 
channels.
This has to be certainly considered as success because there are no 
adjustable parameters in this model calculation.
It is also shown that even a quantitative description of the 
data can be obtained if one introduces small modifications of the
$MN \to \eta N$ amplitudes by exploiting some freedom in the 
$\rho N \to \eta N$ transition potential of the original $MN$ model. 

In our model the dominant role in $\eta$-meson production near 
threshold belongs to the
re-scattering mechanism with intermediate pion exchange. 
The contributions from other meson exchanges, 
specifically from $\rho$ and $\eta$ as well as from the 
direct $\eta$ production are smaller. However, these mechanisms are
still important and have an influence on the cross sections 
due to their interference with the amplitude corresponding to the 
$\pi$ exchange. 

Our study shows that ISI as well as FSI effects are not universal but  
depend, among others, on the concrete meson production mechanisms. 
Thus, a consistent treatment of these effects is very important 
for a quantitative comparison of model calculations with available 
experimental data on the considered reactions.
Specifically, the interaction in the initial $NN$ system 
plays a crucial role since it affects the magnitude of the cross sections,
leading to a significant reduction. 

The presented model should be viewed as a first step towards a consistent 
description of $\eta$ production in $NN$ collisions and in meson induced 
reactions. Future work should avoid some of the technical approximations,
e.g. by an 
explicit treatment of the three--body singularities \cite{Motzke}.
In addition, and most importantly, further mechanisms for $\eta$-meson
production should be explored. 
As mentioned, $\eta$-meson production via $\omega$ rescattering is
still missing in the present model but should be taken into account. 
Furthermore, rescattering contributions
involving the $\Delta$ isobar in the two baryon intermediate 
states \cite{Han2,Han3} should be investigated.
Finally, higher partial waves should be included in order to 
facilitate an extension of the model calculation to higher excess energies
and also to make a comparison with measured angular distributions,    
polarizations and invariant mass spectra meaningful. 

\section*{Acknowledgements}
We would like to thank P. Moskal and K. Nakayama for fruitful discussions. 
This work was partly supported by the DFG-RFBI grant No. 02-02-04001
(436 RUS 113/652/1-1) and by the RFBR grant No. 00-15-96562.

\begin{table}
\begin{center}
\caption{Meson masses, coupling constants and 
cutoff masses utilized in the calculation. The values are
taken from the cited references. Monopole type form factors are 
used at all mesons-baryon vertices.}
\vskip 0.2cm 
\begin{tabular}{|c|c|c|c|c|}
& $m_M$ (MeV) & $\frac{g_{MNN}^2}{4\pi}$ & $\Lambda_{MNN}$ (MeV) & Ref.\\
\hline
 $\pi$ & 138.03 & 14.4 & 1300 &\protect\cite{MHE}\\
 $\rho$ & 769 & 0.84 $(\kappa_{\rho}=6.1)$ & 1400 &\protect\cite{MHE}\\
 $\eta$ & 547.45 & 1.8  & 1500 &\protect\cite{Olipap}\\
 $\sigma$ & 550  & 5.689  & 1700 &\protect\cite{MHE}\\
\end{tabular}
\label{paramet}
\end{center}
\end{table}

\begin{table}[h]
\begin{center}
\caption{Contributions of individual meson exchanges to the $NN\to NN\eta$
cross section and the influence of the $NN$ initial state interaction. 
The values correspond to the excess energy $Q$ = 35 MeV. 
The column labelled 'ratio' exemplifies the actual reduction of the 
cross section after inclusion of the ISI. 
For comparison we also present 
phenomenological reduction factors employed in the model calculations of 
Nakayama et al. \protect\cite{Nakayama} and F\"aldt and Wilkin 
\protect\cite{Wilnew}.
In the column 'CCF' we give the phenomenological reduction factor that 
follows from Eq.~(\ref{isi2}) for the employed $NN$ initial state interaction.}
\vskip 0.2cm 
\begin{tabular}{|c|c|c|c||c|c|c|}
 & \multicolumn{3}{c||}{$\sigma_{NN \to NN\eta}$ [$\mu b$]}   
 & \multicolumn{3}{c|}{phenomenological}\\
 & \multicolumn{3}{c||}{}   
 & \multicolumn{3}{c|}{reduction factors}\\
\hline
$pp\to pp\eta$
& with ISI & without ISI& ratio & CCF & Ref. \cite{Nakayama} & Ref. \cite{Wilnew}\\
\hline
$\pi$    & 3.11 & 9.35 & 0.33 &\  \ $\uparrow$   &\  \ $\uparrow$   &\  \ $\uparrow$   \\
$\rho$   & 0.25 & 10.3 & 0.02 &  &     & \\
$\eta$   & 0.24 & 3.01 & 0.08 &\  \ 0.13&\  \ 0.19 &  \ 0.59  \\
$\sigma$ & 0.36 & 3.17 & 0.11 &  &     &  \\
full result & 5.69 & 10.4 & 0.55 &\  \ $\downarrow$&\  \ $\downarrow$&\  \ $\downarrow$\\
\hline
$pn\to pn\eta$ (I=0) 
& with ISI& without ISI& ratio & CCF & Ref. \cite{Nakayama} & Ref. \cite{Wilnew}\\
\hline
$\pi$    & 10.4& 26.12& 0.4  &\  \ $\uparrow$  &\  \ $\uparrow$  &\  \ $\uparrow$   \\
$\rho$   & 4.30& 25.96& 0.17  & & & \\
$\eta$   & 0.16& 0.85& 0.19 &\  \ 0.30&\  \ 0.27&\  \ 0.53 \\
$\sigma$ & 0.27& 1.05& 0.26 &  &  & \\
full result & 24.48& 70.39&0.35&\  \ $\downarrow$&\  \ $\downarrow$ &\  \ $\downarrow$\\
\end{tabular}
\label{contribISI}
\end{center}
\end{table}

\begin{table}
\begin{center}
\caption{Illustration of interference effects between the reaction amplitudes
based on the $\pi$ and $\rho$ meson-exchange contributions. 
The values correspond to the excess energy $Q$ = 35 MeV.} 
\vskip 0.2cm 
\begin{tabular}{|c|c|c|}
& Original model & Extended model\\
\hline
&  $\frac{|M_{\pi} + M_{\rho}|^2}{|M_{\pi}|^2+|M_{\rho}|^2}$ &
   $\frac{|M_{\pi} + M_{\rho}|^2}{|M_{\pi}|^2+|M_{\rho}|^2}$\\
\hline
I=1 & 0.88  & 0.66\\
\hline
I=0 & 1.13  & 1.45\\
\end{tabular}
\label{pirhocontr}
\end{center}
\end{table}

\vspace*{2.5cm}

\newpage


\end{document}